\documentstyle[epsf,twoside,psfrag,a4,11pt]{article}

\newcommand{\del}{\partial}
\newcommand{\tr}{\mbox{tr}}

\newcommand{\nn}{\nonumber}
\newcommand{\vs}{\vspace*{2ex}}
\newcommand{\Vs}[1]{\vspace*{#1ex}}
\newcommand{\svs}{\vspace*{1ex}}
\newcommand{\mvs}{\vspace*{-2ex}}

\makeatletter
  
  \@addtoreset{equation}{section}
\makeatother

\newcommand{\eq}[1]{(\ref{eq:#1})}

\newcommand{\A}{\!\! & \!\!}
\newcommand{\dal}{\mbox{\raisebox{-.2ex}{\large\boldmath$\Box$}}}

\begin{document}
\renewcommand{\thefootnote}{\fnsymbol{footnote}}
\begin{titlepage}

\begin{flushright}
OU-HET 367\\
hep-th/0010286
\end{flushright}

\Vs{14}
\begin{center}
{\LARGE\bf%
The M2-brane Solution of Heterotic M-theory\\
with the Gauss-Bonnet {\boldmath $R^2$} terms
}\\

\Vs{7}
{\Large%
Ken Kashima\footnote{kashima@het.phys.sci.osaka-u.ac.jp}
}\\

\Vs{2}
{\large\it%
Department of Physics, Osaka University, Toyonaka, %
Osaka 560-0043, Japan
}\\

\end{center}

\Vs{10}
\centerline{{\bf{Abstract}}}
\Vs{-2}
\begin{quotation}
We consider the effective action of 
M-theory compactified on a $S^1/Z_2$
orbifold with $R^2$ interaction in the Gauss-Bonnet combination.
We derive equations of motion with source terms arising from 
the Gauss-Bonnet terms 
and find the M2-brane solution up to order $\kappa^{2/3}$.
It receives a correction
which depends on 
the orbifold coordinate in the same form as the gauge 5-brane solution.
\end{quotation}


\end{titlepage}
\renewcommand{\thefootnote}{\arabic{footnote}}
\setcounter{footnote}{0}
\newpage

\section{Introduction}

Several years ago, Ho\v{r}ava and Witten \cite{HW1,HW2} showed that
the strong coupling limit of ten-dimensional $E_8\times E_8$
heterotic string theory is described by 
M-theory compactified on $M^{10}\times S^1/Z_2$
with a set of $E_8$ gauge fields on two ten-dimensional orbifold 
fixed planes.
The low-energy effective action of this heterotic M-theory consists of
two parts,
$S_{\mbox{\scriptsize SG}}$ and $S_{\mbox{\scriptsize YM}}$.
~$S_{\mbox{\scriptsize SG}}$ is the action of 
usual eleven-dimensional supergravity in the bulk, while
$S_{\mbox{\scriptsize YM}}$ is that of 
super Yang-Mills theory with $E_8$ gauge fields on the orbifold planes.
It is significant to investigate a classical solution of the effective action
for a background of this theory.
Many interesting aspects are discussed
such as the low-dimensional theory with the Calabi-Yau compactification
\cite{W1,O2}, the gaugino condensation \cite{H,O4} and 
various new models \cite{N,RS}.

This effective action is given by an expansion in 
eleven-dimensional gravitational constant $\kappa$ \cite{HW2}.
$S_{\mbox{\scriptsize SG}}$ is zero-th order 
and $S_{\mbox{\scriptsize YM}}$ is first order in $\kappa^{2/3}$.
It is known that, at higher order in $\kappa$,
we need additional interactions of higher powers of
the gauge field $F$ and the curvature $R$ \cite{HW1,HW2,O3}.
In particular, there is $R^2$ interaction in the Gauss-Bonnet combination
at order $\kappa^{2/3}$ \cite{O2,O3}.
It is required by anomaly cancellation and supersymmetry 
as the analogue of ten-dimensional theory \cite{RW}.
In many cases, however, a contribution of the Gauss-Bonnet $R^2$ terms 
is neglected because it is higher order in derivatives.

In the effective theory without the Gauss-Bonnet terms,
the soliton solutions were discussed by Lalak et al.\cite{O1}
The gauge 5-brane solution was constructed explicitly.
This solution has a non-trivial dependence on the orbifold coordinate
because of source terms 
which consist of $E_8$ gauge fields at order $\kappa^{2/3}$.
The $x^{11}$-dependent part of the solution is 
regarded as a correction 
from the strongly coupled heterotic string theory at low-energy limit.
The M2-brane and M5-brane solutions are also discussed.
These solutions, however, do not receive corrections, in contrast to 
the gauge 5-brane solution, because source terms vanish by the brane ansatz.
In addition, it was shown that they are BPS solutions preserving
a quarter of the eleven-dimensional supersymmetry for
the M2-brane oriented orthogonal to  the orbifold planes
as well as the M5-brane oriented parallel to the orbifold planes.
It is consistent with the suggestion \cite{HW1} that
the M2-brane wrapping on $S^1/Z_2$
represents the strongly coupled fundamental heterotic string.

In this paper,
we take the Gauss-Bonnet terms into account.
Since they consist of the metric,
a new contribution to source terms appears in the Einstein equation
and it can not be ignored.
The effect of the Gauss-Bonnet terms is investigated 
in some five-dimensional models \cite{new}, however
there are few discussions on heterotic M-theory.
We expect that these investigations
can reveal new aspects of the low-dimensional models.
To this aim,
we consider the M2-brane solution of heterotic M-theory
and show that it receives a correction of order $\kappa^{2/3}$
from the Gauss-Bonnet terms.

In what follows, 
equations of motion are derived with the Gauss-Bonnet terms.
They are
solved in two asymptotic regions up to order $\kappa^{2/3}$
by an ansatz based on the usual M2-brane solution \cite{DS}.
We show that 
it receives a modification of order $\kappa^{2/3}$
which depends on 
the orbifold coordinate in the same form as the gauge 5-brane case 
\cite{O1}.
This modification can be regarded as a gravitational effect of 
$Z_2$ singularities from the viewpoint of the eleven-dimensional theory,
or the strong coupling correction from the viewpoint of
ten-dimensional string theory.
Finally we discuss interpretations of the solution.
In appendix, we plot the solution numerically 
as a function of $x^{11}$ and $r$,
the coordinates 
in the orbifold direction
and in the transverse direction to M2-brane, respectively.
It confirms that the asymptotic solutions are connected smoothly
and they describe a smooth solution of the field equations.

\newpage

\section{Heterotic M-theory}

We start with the low-energy effective action of 
M-theory on $M^{10}\times S^1/Z_2$, 
namely heterotic M-theory \cite{HW1,HW2},
with the Gauss-Bonnet $R^2$ terms \cite{O2,O3}.
The bosonic part of the action is given by
\begin{eqnarray}
  S= S_{\mbox{\scriptsize SG}}+S_{\mbox{\scriptsize YM}}  
\; .
\end{eqnarray}
Here $S_{\mbox{\scriptsize SG}}$ is the action of familiar
eleven-dimensional supergravity \cite{CJS}
given by
\begin{eqnarray}
 S_{\mbox{\scriptsize SG}} \A=\A \frac{1}{2\kappa^2} \int_{M^{11}} 
  d^{11}x  \biggl\{ \sqrt{-g} 
  \Bigl( -R -\frac{1}{24} G_{IJKL}G^{IJKL} \Bigr) 
  - \frac{\sqrt{2}}{1728}\, \epsilon^{I_1\cdots I_{11}}
   C_{I_1 I_2 I_3}G_{I_4 \cdots I_7}G_{I_8\cdots I_{11}} \biggr\}
\end{eqnarray} 
where $\kappa$ denotes the eleven-dimensional gravitational constant and
$C_{IJK}$ denotes an abelian three-form gauge field 
whose field strength is a four-form given by
$  G_{IJKL} \equiv 4!\,\del_{[I} C_{JKL]}  $.
The indices of $I,J$ denote 
the eleven-dimensional coordinates with $x^0,\cdots , x^9,x^{11}$.
On the other hand, 
$S_{\mbox{\scriptsize YM}}$ describes 
super Yang-Mills theory
on two ten-dimensional orbifold planes 
denoted by $M^{10}_{(1)}$ and $ M^{10}_{(2)}$
for $x^{11}= 0$ and $x^{11}= \pi\rho$,
where we choose $x^{11}$ as the orbifold direction with the range
$x^{11} \in [-\pi\rho, \pi\rho]$.
It is given by\footnote{%
In \cite{A,C}, it is argued that we should multiply
both $S_{\mbox{\tiny YM}}$ and the Bianchi identity \eq{bianchi}
by an additional factor of $2^{-1/3}$.
But we take the original form of \cite{HW2} for simplicity
since this difference is not essential in the following discussion.
}
\begin{eqnarray}
\label{eq:YM}
S_{\mbox{\scriptsize YM}} \A=\A -\frac{1}{8\pi \kappa^2} 
  \left( \frac{\kappa}{4\pi} \right )^{2/3}
  \int_{M^{10}_{(1)} } d^{10}x\, \sqrt{-g}\, 
   \tr \left( F^{(1)}_{AB} F^{(1)AB} \right ) \nn\\
\A\A
   -\frac{1}{8\pi \kappa^2} \left( \frac{\kappa}{4\pi} \right )^{2/3}
   \int_{M^{10}_{(2)}} d^{10}x\, \sqrt{-g} \, 
    \tr \left( F^{(2)}_{AB} F^{(2)AB} \right )
                   \nn \\
 \A\A -\frac{1}{16\pi \kappa^2}\left ( \frac{\kappa}{4\pi} \right )^{2/3}
   \int_{M^{10}_{(1)},\, M^{10}_{(2)}} d^{10}x \, \sqrt{-g} \left (
   R_{ABCD}R^{ABCD}-4\, R_{AB}R^{AB} + R^2 \right )   
\end{eqnarray}
where $F^{(1),(2)}_{AB}$ denote
$E_8$ gauge fields living on the orbifold planes.
The metric in $S_{\mbox{\scriptsize YM}}$ 
is the ten-dimensional part $g_{AB}$ of the eleven-dimensional metric
with the ten-dimensional indices $A,B=0,\cdots,9$.
The Gauss-Bonnet $R^2$ terms appear in the last line in \eq{YM}.

If we consider M-theory on a smooth manifold, 
there is no anomaly. In contrast, on a $S^1/Z_2$ orbifold, there are
gauge and gravitational anomalies owing to $Z_2$ singularities
\cite{AW,HW1,HW2}.
These anomalies can be canceled \cite{HW1,HW2} by the Green-Schwarz
mechanism \cite{GS} with the $E_8$ gauge fields.
At higher order in $\kappa$,
additional interactions are 
required by the anomaly cancellation in terms of higher powers of $F$ and $R$
\cite{HW1,HW2,O3}.
These interactions are investigated 
from a one-loop effect in type IIA string theory \cite{VW}
or anomaly cancellation on world-volume of M5-brane in eleven dimensions
\cite{DLM,W3,A,C}. 
When we reduce this effective action to ten dimensions,
the four-form field strength with the $x^{11}$ index $G_{ABC11}$ 
is promoted to a three-form field strength $H_{ABC}$.
It includes the Yang-Mills and Lorentz Chern-Simons three-forms
due to a modification of the Bianchi identity \eq{bianchi}
by the Green-Schwarz mechanism \cite{HW2,O3}.
In addition, the Gauss-Bonnet $R^2$ terms 
are required by ten-dimensional supersymmetry pairing with
the Lorentz Chern-Simons three-form \cite{RW}.
In the following, 
we concentrate on the effective action up to 
order $\kappa^{2/3}$.

The Einstein equation and the Maxwell equation are given as 
\begin{eqnarray}
\label{eq:einstein}
  R_{IJ} - \frac{1}{2}\, g_{IJ}\, R \A=\A - \frac{1}{24} \left (
  4\, G_{IKLM}G_J{}^{KLM} - \frac{1}{2}\, g_{IJ} \, G_{KLMN}G^{KLMN} \right )
     \nn \\
 \A\A\; -\frac{1}{2\pi} \left( \frac{\kappa}{4\pi} \right )^{2/3} \left\{
  \delta(x^{11})\, T^{(1)}_{IJ} + \delta(x^{11}-\pi\rho)\, T^{(2)}_{IJ}
   \right\}  
\;,
\end{eqnarray}
\begin{eqnarray}
\label{eq:maxwell}
  \del_I \left( \sqrt{-g}\, G^{IJKL}\right) 
     =   \frac{\sqrt{2}}{1152} \, 
  \epsilon^{JKLI_1 \cdots I_8}\, G_{I_1 \cdots I_4} G_{I_5 \cdots I_8}
\end{eqnarray}
where
\begin{eqnarray}
\label{eq:T}
 T^{(i)}_{AB} \A=\A (g_{11,11})^{-1/2}
  \left\{ \tr \left( F^{(i)}_{AC} F^{(i)}_{B}{}^C \right) 
          -\frac{1}{4}\, g_{AB}\, 
             \tr \left( F^{(i)}_{CD} F^{(i)CD} \right)
          + \frac{1}{4}\,
   {\cal G}_{AB}
  \right\}\;, \\
 T^{(i)}_{11,11}\! \A=\A 0 \;.
\end{eqnarray}
We have defined ${\cal G}_{AB}$ in \eq{T} 
as a variation of the Gauss-Bonnet terms 
with respect to the metric:
\begin{eqnarray}
\delta\Bigl\{
   \sqrt{-g}\,( R_{ABCD}R^{ABCD} -4\, R_{AB}R^{AB} + R^2 )
        \Bigr\}
  \equiv \sqrt{-g}\: {\cal G}_{AB}\: \delta g^{AB}  
\end{eqnarray}
where
\begin{eqnarray}
\label{eq:GB}
{\cal G}_{AB}
\A=\A
  - \frac{1}{2}\, g_{AB} 
   ( R_{CDEF}R^{CDEF} -4\, R_{CD}R^{CD}  + R^2 ) \nn\\
 \A\A\; + 2\, R\, R_{AB} + 2\, R_{ACDE}R_B{}^{CDE} 
      -4\, R_{ACBD}R^{CD} -4\, R_A{}^C R_{BC} 
\; .  
\end{eqnarray}
We note that the Einstein equation has source terms 
proportional to $\kappa^{2/3}$. 
The Bianchi identity also has source terms due to a modification \cite{HW2}
for anomaly cancellation and supersymmetry by the Green-Schwarz mechanism:
\begin{eqnarray}
\label{eq:bianchi}
  ( dG)_{11ABCD} = -\frac{1}{2\sqrt{2}\, \pi} 
     \left( \frac{\kappa}{4\pi} \right )^{2/3}  \left \{
  J^{(1)}_{ABCD}\, \delta(x^{11}) + J^{(2)}_{ABCD}\, \delta(x^{11}-\pi\rho)
    \right \}
\end{eqnarray}
where
\begin{eqnarray}
 J^{(i)}_{ABCD} \A=\A 6 
  \left\{
    \tr \left( F^{(i)}_{[AB} F^{(i)}_{CD]} \right)
    -\frac{1}{2}\, \tr \left( R_{[AB}R_{CD]} \right) 
   \right \}
  \; = \; \left( d\, \omega^{(i)}_3 \right)_{ABCD}
\;,
\end{eqnarray}
and $
\omega^{(i)}_3 \equiv \omega^{(i) \mbox{\scriptsize YM} }_3 
   - \frac{1}{2}\,\omega^{ \mbox{\scriptsize L} }_3 
$
denote
the Yang-Mills and Lorentz Chern-Simons three-forms.
$R$ in the trace is the curvature two-form and
the trace is taken over the tangent space indices.

The above description of the Bianchi identity \eq{bianchi}
is called the \lq\lq {\it upstairs picture}''\cite{HW2}
in which we consider the $x^{11}$ coordinate as a circle 
with two $Z_2$ singularities, namely ten-dimensional orbifold planes
at $x^{11}= 0,\pi\rho$.
On the other hand, there is another description 
called the \lq\lq {\it downstairs picture}''
in which the $x^{11}$ coordinate is regarded as a segment with
two boundaries at $x^{11}=0, \pi\rho$. 
In this picture, the Bianchi identity is rewritten as
\begin{eqnarray}
  (dG)_{KLMNP} = 0 
\end{eqnarray}
with the boundary conditions
\begin{eqnarray}
G_{ABCD} \Bigl |_{x^{11}=0} \A = \A
  -\frac{1}{4\sqrt{2}\, \pi} 
     \left( \frac{\kappa}{4\pi} \right )^{2/3}  J^{(1)}_{ABCD}
\;, \\
G_{ABCD} \Bigl |_{x^{11}=\pi\rho} \A = \A
   \frac{1}{4\sqrt{2}\, \pi} 
     \left( \frac{\kappa}{4\pi} \right )^{2/3}  J^{(2)}_{ABCD} 
\end{eqnarray}
where the additional factor of $1/2$ comes from changing a range
of the $x^{11}$ coordinate such that
$ \int^{\:\pi\rho}_{-\pi\rho} dx^{11} 
\rightarrow  2 \int^{\pi\rho}_0 dx^{11}$. 

As mentioned above, equations of motion and the Bianchi identity
have source terms with 
$\delta(x^{11})$. 
This is the reason why the solution of the equations gets a non-trivial 
$x^{11}$-dependence
and it is regarded as the strong coupling effect 
from the point of view of ten-dimensional string theory.
Neglecting the Gauss-Bonnet terms,
the gauge 5-brane solution with the $x^{11}$-dependence 
was computed explicitly by Lalak et al.\cite{O1}
In many cases,
a contribution of the Gauss-Bonnet terms is neglected
since it is higher order in derivatives. 
On the other hand, the $\tr R^2$ terms 
in the modified Bianchi identity \eq{bianchi}
have played the significant role in this effective theory.
For the Calabi-Yau compactification, for instance,
the connection embedding \cite{W1,O2} with 
the $\tr R^2$ terms is very important.
So, to discuss up to order $\kappa^{2/3}$,
the Gauss-Bonnet terms should be considered like 
the $\tr R^2$ terms,
because they are of the same order.
The main aim of this paper is to investigate
the effect of the Gauss-Bonnet terms in heterotic M-theory
at low-energy.

\section{The M2-brane solution}

We consider the M2-brane solution 
up to order $\kappa^{2/3}$.
There are two ways to embed this solution into 
heterotic M-theory depending on whether M2-brane is oriented 
orthogonal or parallel to the orbifold planes.
It has been shown \cite{O1} that, 
neglecting the Gauss-Bonnet terms,
the M2-brane ansatz makes the source terms vanish
and we find the usual solution of eleven-dimensional supergravity \cite{DS}.
In addition, for the $Z_2$ operation of the orbifold,
the orthogonal M2-brane solution is a BPS solution preserving
a quarter of the eleven-dimensional supersymmetry,
while the parallel one does not preserve any supersymmetry.
Then, the orthogonal M2-brane solution
is expected to describe strongly coupled heterotic string effectively. 
On the other hand, 
if we take the the Gauss-Bonnet terms into account, 
they give a new contribution to source terms in the Einstein equation
\eq{einstein} with the non-trivial metric as we have seen in 
\eq{T} and \eq{GB}.
In the following,
we consider the usual orthogonal M2-brane ansatz
but we will show that it receives a correction of order $\kappa^{2/3}$.

\subsection{The ansatz and the linearization of equations}

\psfrag{xM}{$x^M$}
\psfrag{xm}{$x^m$}
\psfrag{xa}{$x^\alpha$} 
\psfrag{x11}{$x^{11}${\footnotesize: the orbifold direction}}
\psfrag{x0}{$\scriptstyle x^{11}=0$}
\psfrag{xA}{$x^A$}
\psfrag{xpi}{$\scriptstyle x^{11}=\pi\rho$}
\psfrag{etc}{
$
\begin{array}{rcl}
 M\A=\A 0,\cdots,9,11 \\
 A\A=\A0,\cdots,9  \\
 \mu\A=\A0,1,11 \\
 \alpha\A=\A0,1 \\
 m\A=\A 2,\cdots,9
\end{array}
$
}

\begin{figure}
\epsfysize = 8cm
\centerline{\epsfbox{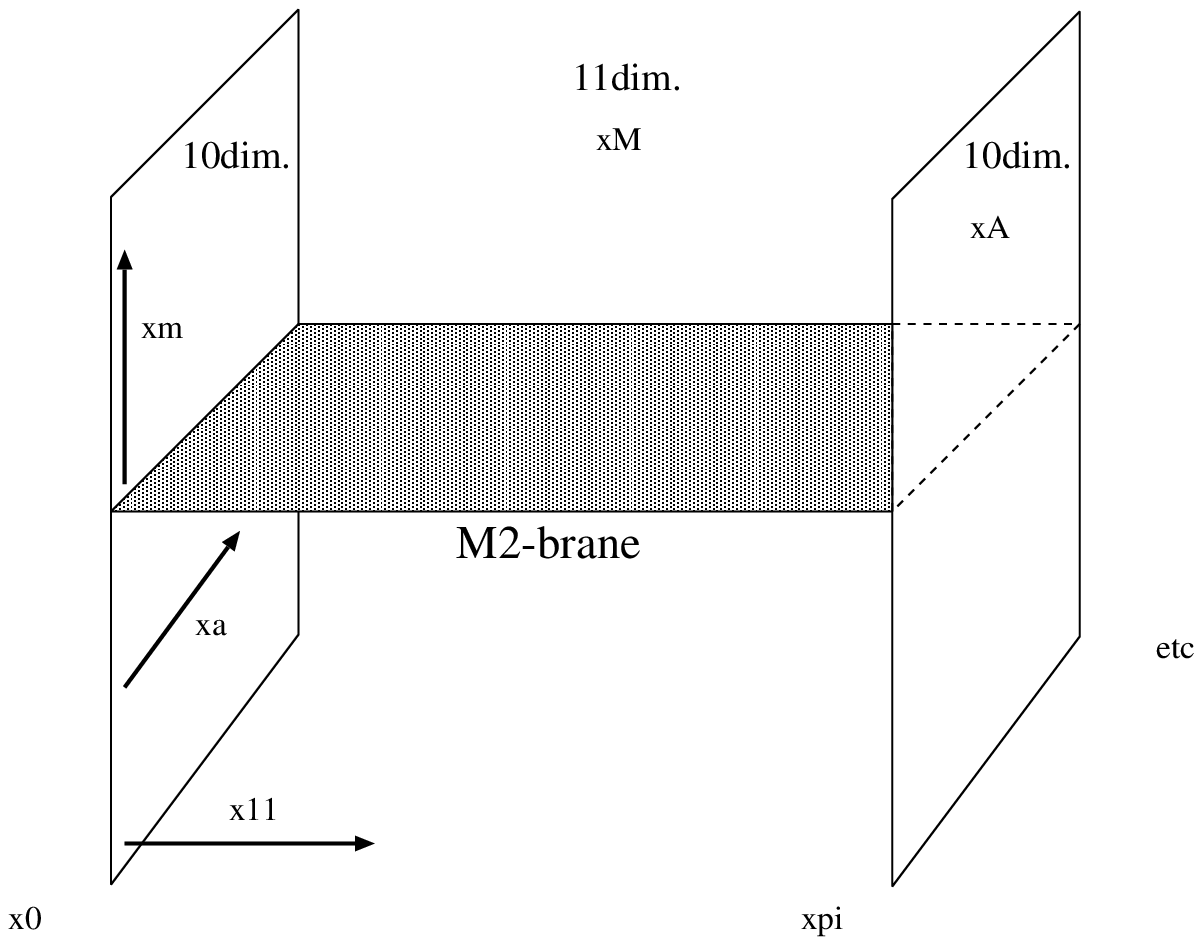}}
\caption{Indices of coordinates}
\end{figure}

We split the eleven-dimensional coordinates $x^M$ into
$x^{\mu}\; (\mu=0,1,11)$ and $x^m\; (m=2,3,\cdots ,9)$ 
for the world-volume coordinates of M2-brane and
the transverse coordinates respectively.
Furthermore, we split $x^{\mu}$ into $x^{\alpha}\;(\alpha = 0,1)$ and $x^{11}$
where $x^{\alpha}$ denotes the coordinates of projected M2-brane
onto the orbifold planes, as summarized in Figure 1.
We take an ansatz for the metric in the following form:
\begin{eqnarray}
\label{eq:metric}
 g_{MN} = 
  \left( 
   \begin{array}{ccc}
     g_{\alpha\beta} & & 0 \\
      & g_{mn} & \\ 
      0 &&  g_{11,11}
   \end{array}
            \right )
=
   \left( 
   \begin{array}{ccc}
     e^{2A}\, \eta_{\alpha\beta} & & 0 \\
      & e^{2B}\, \delta_{mn} & \\ 
      0 &&  e^{2X}
   \end{array}
            \right )
\end{eqnarray}
with
\begin{eqnarray}
\label{eq:warp1}
 A \A=\A \frac{1}{3}\, Y_0 + \kappa^{2/3}\Phi_{1A}
         + \mbox{O}(\kappa^{4/3}) \;,  \nn \\
 B \A=\A -\frac{1}{6}\, Y_0 + \kappa^{2/3} \Phi_{1B}
         + \mbox{O}(\kappa^{4/3}) \;,  \\
 X \A=\A \frac{1}{3}\, Y_0 + \kappa^{2/3}\Phi_{1X}
         + \mbox{O}(\kappa^{4/3})  \nn 
\end{eqnarray}
where
the subscripts $\lq\lq\, 0$ '' and $\lq\lq\,1$ '' denote 
zero-th and first order in $\kappa^{2/3}$ respectively.
The zero-th order term $Y_0$ is a known function defined later in \eq{y0}.
At order $\kappa^0$, this metric 
describes orthogonal M2-brane which spreads on the directions of
$x^0, x^1$ and $x^{11}$ at $x^m=0$.
For the three-form $C_{MNL}$, we require 
$ C_{\mu\nu\rho}   \equiv  \frac{1}{\sqrt{2}\,3!}\,\epsilon_{\mu\nu\rho}
  \, e^C  $
and set others to zero
with the field strength:
\begin{eqnarray}
\label{eq:G}
    G_{m \mu\nu\rho} =  \frac{1}{\sqrt{2}}\,\epsilon_{\mu\nu\rho}
  \,\del_m \, e^C  \; , \quad
   \mbox{others} =  0  
\end{eqnarray}
where
\begin{eqnarray}
\label{eq:warp2}
   C = Y_0 + \kappa^{2/3}\Phi_{1C}
         + \mbox{O}(\kappa^{4/3})  
\; .
\end{eqnarray}
We note that, 
at order $\kappa^0$,  the ansatz leads to the usual M2-brane solution
\cite{DS}
with $  A = X = 1/3\, C$ and $ B = -1/6\, C $.
At order $\kappa^{2/3}$, fields depend on $x^{11}$ as well as $r$ 
due to the source terms with $\delta(x^{11})$ in equations of motion 
\eq{einstein}.
Then $\Phi_{1A}, \Phi_{1B}, \Phi_{1X}$ and $\Phi_{1C}$ are functions of
$x^{11}$ and $r$. 
We turn off the $E_8$ gauge fields to see the effect of the pure Gauss-Bonnet
terms on the M2-brane solution.

\vs
Now, we linearize equations of motion by the ansatz, \eq{metric} and \eq{G}.
Substituting the ansatz to the Einstein equation \eq{einstein}, 
we find the following conditions for $(\alpha,\beta), (m,n)$ and $(11,11)$ 
components of indices:
\begin{eqnarray}
 \label{eq:e-a}
\A (\alpha,\beta)\,: \A  \qquad\qquad
  \frac{1}{2}\, e^{Y_0} \dal e^{-Y_0}  \nn \\
\A\A \qquad\qquad 
   \; + \kappa^{2/3}  \Bigl[
     \del Y_0\,\del Y_0 ( -\Phi_{1A} -\frac{1}{2}\,\Phi_{1X} 
    +\frac{1}{2}\,\Phi_{1C} )     
  + \del Y_0 ( -3\del \Phi_{1B} + \frac{1}{2}\, \del \Phi_{1C} ) \nn \\
\A\A \qquad\qquad
  \qquad \qquad   + \dal\Phi_{1A} +7\dal \Phi_{1B}  + \dal \Phi_{1X} 
   + {e^{-Y_0}}\, \del_{11}^2( \phi_A +8\phi_B ) \Bigr]  \nn\\
\A\A  \qquad\qquad
  \;
    = \frac{1}{2} \, \kappa^{2/3}
    \left\{ \delta(x^{11})+\delta(x^{11}-\pi\rho)\right\}
    J_1  + \mbox{O}(\kappa^{4/3})  \;, \\
 \nn \\
 \label{eq:e-m}
\A (m,n)\,: \A \qquad\qquad
 \kappa^{2/3}  \Bigl[
  (  \del Y_0\, \del Y_0 + \del Y_0\,\del) 
( 6\Phi_{1A} + 3\Phi_{1X} -3\Phi_{1C} )   \nn\\
\A\A \qquad\qquad 
  \qquad  \quad
    + 14 \dal\Phi_{1A} +42\dal\Phi_{1B}+7\dal\Phi_{1X} 
   + {e^{-Y_0}}\, \del_{11}^2( 16\phi_A + 56\phi_B) \Bigr] \nn \\
\A\A \qquad\qquad
  \;  = \kappa^{2/3} \left\{ \delta(x^{11})+\delta(x^{11}-\pi\rho)\right\}
    J_2   + \mbox{O}(\kappa^{4/3}) \;, \\ 
\nn \\
\A(11, 11)\,:\A \qquad\qquad
 \label{eq:e-11}
  \frac{1}{2}\, {e^{Y_0}} \dal {e^{-Y_0}}  \nn \\
\A\A \qquad\qquad
 \; + \kappa^{2/3}  \Bigl[
    \del Y_0\,\del Y_0 ( -\Phi_{1A} -\frac{1}{2}\,\Phi_{1X} 
    +\frac{1}{2}\,\Phi_{1C} )     
  + \del Y_0 ( -3\del \Phi_{1B} + \frac{1}{2}\, \del \Phi_{1C} ) \nn \\
\A\A \qquad\qquad
  \qquad\qquad + 2\dal\Phi_{1A} +7\dal \Phi_{1B} \Bigr]
     + \mbox{O}(\kappa^{4/3})  = 0
\end{eqnarray}
where  
$\del$ denotes the derivative with respect to $x^m$ and
$\dal$ denotes $\delta^{mn}\del_m\del_n$.
We define $J_1$ and $J_2$ as contributions of $T^{(i)}$ in \eq{T}
to the source terms.
Since we turn off $E_8$ gauge fields $F^{(i)}$, 
only the variation ${\cal G}$ of the Gauss-Bonnet terms in \eq{GB} 
appears in $J_1$ and $J_2$.
The Maxwell equation \eq{maxwell} 
leads to two conditions:

\svs
\noindent
$(\nabla_m G^{\mu\nu\rho m})$ :
\begin{eqnarray}
 \label{eq:m-1} 
  -\dal e^{-Y_0}
  + \kappa^{2/3}\,  e^{-Y_0} \left\{
  \del Y_0 ( -2\del\Phi_{1A}+6\del\Phi_{1B}-\del\Phi_{1X} )
   + \dal \Phi_{1C} 
   \right\}   + \mbox{O}(\kappa^{4/3})   
    = 0  
\;,
\end{eqnarray}
\noindent
$(\nabla_{\mu} G^{\mu\nu\rho m})$ :
\begin{eqnarray}
 \label{eq:m-2}
  \kappa^{2/3}\, e^{-Y_0}
  \bigl\{
   \del Y_0 ( -2\del_{11}\Phi_{1A} + 6\del_{11}\Phi_{1B}-\del_{11}\Phi_{1X} 
    +\del_{11}\Phi_{1C}  )    + \del_{11}\del \Phi_{1C} 
   \bigr\} + \mbox{O}(\kappa^{4/3})   = 0    
\; .
\end{eqnarray}
We consider the Bianchi identity \eq{bianchi}.
Since we turn off the gauge fields,
source terms are given by $\tr R^2$ terms.
However it is straightforward to show \cite{O1} that
the $\tr R^2$ vanishes at order $\kappa^0$ for its anti-symmetric indices
upon substituting the ansatz of the metric \eq{metric}. For this reason,
the Bianchi identity does not lead to any additional condition up to order 
$\kappa^{2/3}$.

At order $\kappa^0$, the above equations of motion \eq{e-a}--\eq{m-2}
lead to the usual 
field equation $\dal e^{-Y_0} = 0$ for eleven-dimensional supergravity. 
The solution is given by
\begin{eqnarray}
\label{eq:y0}
  e^{-Y_0} = 1+ \frac{Q}{r^6}  
\end{eqnarray}
where $Q$ denotes a charge of M2-brane \cite{DS}.

Next, we compute the source terms, $J_1$ and $J_2$, 
in the Einstein equation, \eq{e-a} and \eq{e-m}.
Substituting the metric \eq{metric} to ${\cal G}$ in \eq{GB},
we get 
\begin{eqnarray}
\label{eq:g-a}
g^{\alpha\beta}\,{\cal G}_{\alpha\beta} \A=\A
 e^{\frac{2}{3} Y_0} \Bigl(
    - \frac{2}{3}\,\dal Y_0\,\dal Y_0 
    +\frac{2}{3}\,\del_m\del_n Y_0\, \del_i\del_j Y_0 \,\delta^{mi}\delta^{nj}
    +\frac{1}{3}\,\dal Y_0\,\del_m Y_0\,\del_n Y_0\,\delta^{mn}
\nn\\
\A\A \qquad \quad
    -\frac{10}{9}\,\del_m\del_n Y_0\,\del_i Y_0\,\del_j Y_0\,
                                                         \delta^{mi}\delta^{nj}
    -\frac{7}{54}\,\del_m Y_0\,\del_n Y_0\,\del_i Y_0\, \del_j Y_0\,
                                                        \delta^{mn}\delta^{ij}
    \Bigl)\; +\;\mbox{O}(\kappa^{2/3})
\nn \\
\A=\A 36 Q^2 (r^6 +Q )^{-\frac{14}{3}}  
  \Bigl(  -\frac{14}{3}\,Q^2 +\frac{196}{3}\,Qr^6 + \frac{112}{3}\,r^{12}
  \Bigr) \; +\;\mbox{O}(\kappa^{2/3})
\;,
\end{eqnarray}
\begin{eqnarray}
\label{eq:g-m}
g^{mn}\,{\cal G}_{mn} \A=\A
 e^{\frac{2}{3} Y_0} \Bigl(
    4\,\dal Y_0\,\dal Y_0 
   -4\,\del_k\del_l Y_0\, \del_i\del_j Y_0 \,\delta^{ki}\delta^{lj}
    -\frac{8}{9}\,\dal Y_0\,\del_i Y_0\,\del_j Y_0\,\delta^{ij}
\nn\\
\A\A \qquad \quad
    -\frac{16}{3}\,\del_k\del_l Y_0\,\del_i Y_0\,\del_j Y_0\,
                                                         \delta^{ki}\delta^{lj}
    -\frac{7}{27}\,\del_k Y_0\,\del_l Y_0\,\del_i Y_0\, \del_j Y_0\,
                                                        \delta^{kl}\delta^{ij}
    \Bigl)\; +\;\mbox{O}(\kappa^{2/3})
\nn\\
\A=\A 36 Q^2 (r^6 +Q )^{-\frac{14}{3}}
  \Bigl(  \frac{308}{3}\,Q^2 +112\,Qr^6 -224\,r^{12}
  \Bigr) \; +\;\mbox{O}(\kappa^{2/3})
\;.
\end{eqnarray}
From \eq{einstein}, 
\eq{g-a} and \eq{g-m},
we find
\begin{eqnarray}
\label{eq:j1}
    J_1 \A=\A 36 \alpha \,(r^6+Q)^{-4}\, r^{-4} Q^2 
  \Bigl(  -\frac{14}{3}\,Q^2 +\frac{196}{3}\,Qr^6 + \frac{112}{3}\,r^{12}
  \Bigr)
\;, \\
\label{eq:j2}
  J_2 \A=\A 36 \alpha \,(r^6+Q)^{-4}\, r^{-4} Q^2 
  \Bigl(  \frac{308}{3}\,Q^2 +112\,Qr^6 -224\,r^{12}
  \Bigr)
\end{eqnarray}
where $ \alpha \equiv 1/{8\pi ( 4\pi )^{2/3}} $.

\vs
We obtained equations for $\Phi_{1A}, \Phi_{1B},
\Phi_{1X}$ and $\Phi_{1C}$ at order $\kappa^{2/3}$. 
At this stage,
these four functions $\Phi_1$ are unknown, while we have 
two conditions from the Maxwell equation, \eq{m-1} and \eq{m-2}, 
and three conditions from the Einstein equation \eq{e-a}--\eq{e-11}.
One may wonder there are too many conditions.
But the two conditions from the Maxwell equation 
are related by an arbitrary function $f$, as we will see below. 
Then the number of functions matches to the number of independent
conditions and we can solve these equations.

We separate $\Phi_1$ into $x^{11}$-independent and $x^{11}$-dependent
parts 
\begin{eqnarray}
\label{eq:dcmps}
 \Phi_{1A} \equiv A_1 + \phi_A \;,\quad
 \Phi_{1B} \equiv B_1 + \phi_B \;,\quad
 \Phi_{1X} \equiv X_1 + \phi_X \;,\quad
 \Phi_{1C} \equiv C_1 + \phi_C 
\end{eqnarray}
where $ A_1, B_1, X_1$ and $ C_1 $ depend only on 
$r$, while $ \phi_A, \phi_B, \phi_X$ and $\phi_C$  are functions
of $x^{11}$ and $r$.
To determine these  decompositions uniquely, 
we demand \cite{O1}
\begin{eqnarray}
\label{eq:i-c}
  \int^{\pi\rho}_0 \phi_A\, dx^{11} =  \int^{\pi\rho}_0 \phi_B \,dx^{11}
 = \int^{\pi\rho}_0 \phi_X \,dx^{11} = 0
\;.
\end{eqnarray}
For $\phi_C$, on the other hand, we need a different condition.
If we require the same condition $\int^{\pi\rho}_0 \phi_C \, dx^{11} = 0$ 
as other $\phi$, we obtain
two independent conditions for $\phi_A$ and $\phi_B$ 
from the $(\alpha,\beta)$ and $(m,n)$ components of the Einstein equation,
\eq{e-a} and \eq{e-m}.
Together with the $(11,11)$ component \eq{e-11},
we find three conditions which are too strong for $\phi_A$ and $\phi_B$.
So, there is no solution with $\int^{\pi\rho}_0 \phi_C \, dx^{11} = 0$.
For this reason, we require $\int^{\pi\rho}_0 \phi_C \, dx^{11} \neq 0$
and will determine the value by equations of motion, below.

\vs
Now, let us decompose the equations of motion \eq{e-a}--\eq{m-2}
according to \eq{dcmps} and \eq{i-c}.
For the Maxwell equation, \eq{m-1} and \eq{m-2}, we find that 
\begin{eqnarray}
  \label{eq:m-3}
\A\A  \del Y_0 ( -2\del A_1 + 6\del B_1 - \del X_1 ) + \dal C_1 
 + \del Y_0 ( -2\del\phi_A +6\del\phi_B -\del\phi_X ) 
        + \dal \phi_C =0  \;, \\
\label{eq:m-4}
\A\A
 \del Y_0 ( -2\del_{11}\phi_A +6\del_{11}\phi_B -\del_{11}\phi_X 
   + \del_{11}\phi_C) + \del_{11}\del \phi_C =0  \;.
\end{eqnarray}
For the $x^{11}$-dependent part,
\eq{m-4} is integrated over $x^{11}$ with a result
\begin{eqnarray}
  \label{eq:s-6}
  \del Y_0 (  -2\phi_A + 6\phi_B - \phi_X + \phi_C) + \del \phi_C 
  + \del Y_0 \, f = 0 
\end{eqnarray}
where $f$ is an arbitrary function of $r$ arising from the integration.
Differentiating \eq{s-6} by $x^m$, we have
\begin{eqnarray}
  \label{eq:s-7}
    \del Y_0 (  -2\del\phi_A + 6\del\phi_B - \del\phi_X  ) + 
    \dal \phi_C   + \del Y_0\, \del f = 0 
\; .
\end{eqnarray}
Substituting \eq{s-7} to \eq{m-3}
yields, for the $x^{11}$-independent part,
\begin{eqnarray}
  \label{eq:s-1}
    \del Y_0 (  -2\del A_1 + 6\del B_1 - \del X_1 ) + 
    \dal C_1   - \del Y_0\, \del f = 0 
\end{eqnarray}
and, after integrating over $x^m$, it leads to
\begin{eqnarray}
  \label{eq:s-2}
  \del Y_0 (  -2A_1 + 6 B_1 - X_1 + C_1) + \del C_1 - \del Y_0\,f +g =0 
\end{eqnarray}
where $g$ is a constant of integration.

Next, we turn to the Einstein equation.
For the $(11,11)$ component,
we substitute \eq{s-6} to \eq{e-11} and integrate it over $x^{11}$.
Then, according to \eq{i-c},
we find that
\begin{eqnarray}
  \label{eq:s-5}
\A\A  -3\del Y_0 \del Y_0 \,B_1 -3 \del Y_0 \del B_1 
  + 2\dal A_1 + 7\dal B_1 -\frac{1}{2}\, g =0 \;, \\
  \label{eq:s-10}
\A\A  -3\del Y_0 \del Y_0 \,\phi_B -3 \del Y_0 \del \phi_B
  + 2\dal \phi_A + 7\dal \phi_B = 0 \;.  
\end{eqnarray}
Consider the $(\alpha,\beta)$ component.
Substituting \eq{e-11} to \eq{e-a}, we obtain 
\begin{eqnarray}
  -\dal \Phi_A + \dal \Phi_X + e^{-Y_0}\,\del^2_{11}( \phi_A + 8\phi_B)
 = \frac{1}{2}
 \left\{ \delta(x^{11})+\delta(x^{11}-\pi\rho)\right\}
    J_1 \;.  
\end{eqnarray}
In the \lq\lq {\it downstairs picture}'', this equation is rewritten 
as
\begin{eqnarray}
 \label{eq:s-11}
   - \dal \Phi_A + \dal \Phi_X + e^{-Y_0}\,\del^2_{11}( \phi_A + 8\phi_B)
 = 0 
\end{eqnarray}
with boundary conditions
\begin{eqnarray}
\A\A
 \bigl[ e^{-Y_0}\, \del_{11}( \phi_A + 8\phi_B) \bigr] \Big|_{x^{11}=0}  
 = \frac{1}{4}\, J_1 \;,  \\
\A\A
 \bigl[ e^{-Y_0}\, \del_{11}( \phi_A + 8\phi_B) \bigr] \Big|_{x^{11}=\pi\rho}  
 = - \frac{1}{4}\, J_1  \;. 
\end{eqnarray}
Integrating \eq{s-11} over $x^{11}$, we find 
\begin{eqnarray}
 \label{eq:s-3}
 - \dal A_1 + \dal X_1 = \frac{1}{2}\, J_1 \,\frac{1}{\pi\rho} 
\end{eqnarray}
for the $x^{11}$-independent part,
and 
\begin{eqnarray}
   \label{eq:s-8}
\A\A - \dal \phi_A +\dal \phi_X + e^{-Y_0}\,\del^2_{11}( \phi_A + 8\phi_B)
 = -
 \frac{1}{2}\,J_1 \,\frac{1}{\pi\rho}\;, \nn \\
\A\A \qquad
 \bigl[ e^{-Y_0}\, \del_{11}( \phi_A +8 \phi_B) \bigr] \Big|_{x^{11}=0}  
 = \frac{1}{4}\, J_1  \;,\\
\A\A \qquad
 \bigl[ e^{-Y_0}\, \del_{11}( \phi_A + 8\phi_B) \bigr] \Big|_{x^{11}=\pi\rho}  
 = - \frac{1}{4}\, J_1 \nn 
\end{eqnarray}
for the $x^{11}$-dependent part.
The remaining $(m,n)$ component is also worked out in
the same way as the $(\alpha,\beta)$ component.
In the {\it downstairs picture}, results are
\begin{eqnarray}
  \label{eq:s-4}
 ( \del Y_0\,\del Y_0 + \del Y_0 \,\del)
( 6 A_1 + 3 X_1 -3 C_1 ) + 14\dal A_1
  + 42\dal B_1 +7 \dal X_1 = J_2\,\frac{1}{\pi\rho} 
\end{eqnarray}
and
\begin{eqnarray}
  \label{eq:s-9}
\A\A
   ( \del Y_0 \,\del Y_0 + \del Y_0\,\del)
 ( 6 \phi_A + 3 \phi_X -3 \phi_C ) 
        +14 \dal \phi_A  + 42\dal \phi_B +7 \dal \phi_X \nn \\
\A\A \;
     \; + e^{-Y_0}\,\del^2_{11}( 16\phi_A + 56\phi_B)
       = - J_2\,\frac{1}{\pi\rho} \nn\;, \\
\A\A \quad
     \Bigl[ -3\del Y_0\, \del \int \phi_C\, dx^{11}
     + e^{-Y_0}\,\del_{11}( 16\phi_A + 56\phi_B) \Bigr]
     \Big|_{x^{11}=0} = \frac{1}{2}\, J_2 \;,  \\
\A\A \quad
     \Bigl[ -3\del Y_0\, \del \int \phi_C\, dx^{11}
     + e^{-Y_0}\,\del_{11}( 16\phi_A + 56\phi_B) \Bigr]
     \Big|_{x^{11}=\pi\rho} = - \frac{1}{2}\, J_2  
\nn
\;.
\end{eqnarray}
We have finished decomposing all equations of motion,
which are summarized in the table below.
\begin{center}
\begin{tabular}{|c|c|c|}
\hline
 & The $x^{11}$-independent part & The $x^{11}$-dependent part \\ \hline
The Maxwell equation & \eq{s-1} & \eq{s-7} \\ \hline
The Einstein equation & & \\
$(\alpha\beta)$ & \eq{s-3} & \eq{s-8} \\
$(m,n)$   & \eq{s-4}  & \eq{s-9} \\
$(11,11)$ & \eq{s-5} & \eq{s-10} \\ \hline
\end{tabular}
\end{center}

\subsection{Solving equations at order $\kappa^{2/3}$}

We solve the equations in two asymptotic regions,
$r^6 \gg Q$ and $r^6 \ll Q$, 
by expanding fields in $r$. 
In these regions 
we find asymptotically, 
in $r^6 \gg Q$ :
\begin{eqnarray}
\label{eq:rgq}
   e^{-Y_0} = 1  \;, \quad
   \del^m Y_0 = 6\,Q\, r^{-8} x^m \;, \quad
  J_1 =  j_1\,\alpha Q^2\: r^{-16} \;, \quad 
    J_2 = j_2\,\alpha Q^2\: r^{-16}  
\end{eqnarray}
where 
$     j_1\equiv 1344 $ and $  j_2 \equiv -8064
$ , while in $r^6 \ll Q$ : 
\begin{eqnarray}
\label{eq:rlq}
 e^{-Y_0} = r^{-6} Q  \;,\quad
  \del^m Y_0 = 6\, r^{-2} x^m \;, \quad
  J_1 = \tilde j_1\,\alpha \: r^{-4} \;, \quad 
    J_2 = \tilde j_2\,\alpha \: r^{-4}  
\end{eqnarray}
where
$     \tilde j_1\equiv 
                 -168
$ and $ \tilde j_2 \equiv 
                 3696 
$ .

\subsubsection{$ r^6\gg Q $ region}

First, we consider the $x^{11}$-dependent part.
Since source terms become zero at $ r \rightarrow \infty $ limit by \eq{rgq},
we are interested in a solution which is regular at $ r \rightarrow \infty $
limit.
We should expand fields in $r^{-1}$ as
\begin{eqnarray}
  \phi (r,x^{11}) = \sum_{i=0}^{\infty} \psi_{p+i}\; r^{-(p+i)}
\end{eqnarray}
for a certain integer $p$ where $\psi_i$ is a function of $x^{11}$.
In this expression,
$\psi_p\, r^{-p}$ is dominant and we can regard
$\psi_i\, r^{-i}$ for $i>p$ as a small correction.
If we assume $\phi \sim  r^{-p}$ approximately
without the $x^{11}$-dependence,
then $p$ can be determined by 
source terms, $J_1$ and $J_2$, as follows.
\begin{enumerate}
\item 
From the $(11,11)$ component of the Einstein equation \eq{s-10},
we find $\phi_A \sim \phi_B$.
\item
The $(\alpha,\beta)$ component \eq{s-8} yields 
$\phi_A \sim \phi_B \sim J_1 \sim r^{-16}$.
\item
From boundary conditions of the $(m,n)$ component \eq{s-9},
we obtain 
\mvs
$$ \del Y_0\,\del \! \int \!\phi_C\,dx^{11} \Big|_{x^{11}=0,\pi \rho}
\sim r^{-16} 
\mvs
$$
which tells us $\phi_C \sim r^{-8}$.
\item
We integrate the Maxwell equation \eq{s-7} over $x^{11}$ from 0 to $\pi\rho$
with a result
\mvs
$$\dal \int_0^{\pi\rho} \! \phi_C\, dx^{11} + \del Y_0\, \del f\, \pi\rho=0
\mvs
$$
which amounts to $ f\sim r^{-2}$.
Moreover, substituting the result to \eq{s-7} again, we obtain
$ \phi_X\sim \phi_A\sim \phi_B\sim r^{-16}$.
\end{enumerate}

\noindent
The above discussion is very rough and ignores the $x^{11}$-dependence,
but we find a significant result
\begin{eqnarray}
\label{eq:c-1}
\phi_A = {\displaystyle \sum^{\infty}_{i=0}} 
          \psi^{(A)}_{16+i}\;r^{-16-i}\;,\quad
\phi_B = {\displaystyle \sum^{\infty}_{i=0}} 
         \psi^{(B)}_{16+i}\;r^{-16-i}\;,\quad
\phi_X = {\displaystyle \sum^{\infty}_{i=0}} 
         \psi^{(X)}_{16+i}\;r^{-16-i}\;,
\nn\\
\phi_C = {\displaystyle \sum^{\infty}_{i=0}} 
          \psi^{(C)}_{8+i}\;r^{-8-i}\;,\quad
  f = {\displaystyle\sum^{\infty}_{i=0}} P^{(f)}_{2+i} \;r^{-2-i}  
\end{eqnarray}
where 
$\psi^{(A)}_i, \psi^{(B)}_i, \psi^{(X)}_i $ and $ \psi^{(C)}_i$
are functions of $x^{11}$,
while $P_i^{(f)}$ is a constant.
We note that $\int_0^{\pi\rho} \phi_C\,dx^{11}$ is determined by $f$
as mentioned above.

Next,
we expand the Maxwell equation \eq{s-7} and the Einstein equation, 
\eq{s-10}, \eq{s-8} and \eq{s-9}, in $r$.
According to \eq{c-1}, it is sufficient to consider 
the following parts of equations.

The Maxwell equation:
\begin{eqnarray}
\label{eq:dec-1}
\A\A \quad \sum_i 
   \biggl[ -6(i-8)Q \Bigl\{ -2 \psi^{(A)}_{i-8} +6 \psi^{(B)}_{i-8}
   - \psi^{(X)}_{i-8} \Bigr\} 
\nn \\
\A\A \qquad\quad 
     +(i-2)(i-8)\psi^{(C)}_{i-2}  
     -6(i-8)Q P^{(f)}_{i-8}  \biggr]\: r^{-i}
      = 0 \;, 
\qquad   \underline{ i \geq 10}
\end{eqnarray}
\indent
The Einstein equation
\begin{eqnarray}
\label{eq:dec-2}
\A(\alpha,\beta)\,:\A \quad
\sum_i  \biggl[ 
  (i-2)(i-8)\Bigl\{ -\psi^{(A)}_{i-2} + \psi^{(X)}_{i-2} \Bigr\}
  + \del^2_{11}\psi^{(A)}_i + 8 \del^2_{11}\psi^{(B)}_i
 \biggr]\: r^{-i} 
\nn \\
\A\A  \qquad
 = -\frac{1}{2}\, j_1\,\alpha Q^2\: r^{-16} \,\frac{1}{\pi\rho}
\;, \makebox[2in]{} \underline{  i \geq 16}
 \\
\label{eq:dec-3}
\A(m,n)\,:\A \quad
 {\displaystyle \sum_i }
    \biggl[  
       36 Q^2 \Bigl\{ 6 \psi^{(A)}_{i-14}  
       + 3\psi^{(X)}_{i-14}  -3\psi^{(C)}_{i-14}  \Bigr\} 
\nn\\
\A\A\quad\qquad
       -6(i-8)Q \Bigl\{ 6 \psi^{(A)}_{i-8}  
       + 3\psi^{(X)}_{i-8}  -3\psi^{(C)}_{i-8}  \Bigr\} 
   \nn\\
 \A\A\quad\qquad
       +(i-2)(i-8)\Bigl\{ 14 \psi^{(A)}_{i-2} 
       +42\psi^{(B)}_{i-2}  +7\psi^{(X)}_{i-2}  \Bigr\} 
   \nn \\
 \A\A\quad\qquad 
       + 16\del^2_{11}\psi^{(A)}_i +56\del^2_{11}\psi^{(B)}_i 
    \biggr]\: r^{-i}
    = -j_2\, \alpha Q^2\:r^{-16}\,\frac{1}{\pi\rho} 
\;, \qquad \underline{  i \geq 16}
\\
\label{eq:dec-4}
\A(11,11)\,:\A \quad
 {\displaystyle \sum_i} 
 \biggl[ 
    -108 Q^2 \psi^{(B)}_{i-14} + 18(i-8)Q \psi^{(B)}_{i-8} 
\nn\\
\A\A\quad\qquad
  + (i-2)(i-8)\Bigl\{ 
    2 \psi^{(A)}_{i-2} +7 \psi^{(B)}_{i-2} \Bigr\}
 \biggr] \: r^{-i} 
= 0 
\;, \qquad\qquad \underline{  i \geq 18}
\end{eqnarray}
We note that it is always necessary to keep boundary conditions 
in mind with the $(\alpha,\beta)$ and $(m,n)$ components.

Let us determine $\psi_i$ and $P^{(f)}_i$.
In $r^6\gg Q$ region, it is enough to consider the dominant part:
$  \psi_{16}^{(A)},\;\psi_{16}^{(B)},\;\psi_{16}^{(X)},\;
  \psi_8^{(C)}$ and $P_2^{(f)}$.
$i=18$ part of \eq{dec-4}
gives $
\psi^{(A)}_{16}  =  -7/2\,\psi^{(B)}_{16} $, and then
$i=16$ part of \eq{dec-2} yields
\begin{eqnarray}
  \del^2_{11} \psi_{16}^{(B)}
 = - \frac{1}{9}\,j_1 \,\alpha Q^2\, \frac{1}{\pi\rho} 
\end{eqnarray}
with boundary conditions 
\begin{eqnarray}
 \del_{11} \psi^{(B)}_{16}  \Big|_{x^{11}=0}  
 = \frac{1}{18}\,j_1 \,\alpha Q^2
\;,\quad
 \del_{11} \psi^{(B)}_{16}  \Big|_{x^{11}=\pi\rho}  
 = - \frac{1}{18}\,j_1 \,\alpha Q^2
\; .
\end{eqnarray}
From these conditions, we have
\begin{eqnarray}
\label{eq:sol-1}
  \psi^{(A)}_{16} \A = \A -\frac{7}{2}\,\psi^{(B)}_{16}  
  = 
     \frac{1568}{3} \,
   \alpha Q^2
   \left\{
    \frac{1}{2\pi\rho}(x^{11})^2 - \frac{1}{2}\,x^{11} 
    + \frac{\pi\rho}{12}
   \right\}  
\;.
\end{eqnarray}
From  $i=16$ part of \eq{dec-3} and $i=10$ part of \eq{dec-1},
~$\psi^{(C)}_8$ and $P_2^{(f)}$ are given by
\begin{eqnarray}
\label{eq:sol-2}
 \psi^{(C)}_8 
    \A =\A
                56
            \,\alpha Q\,\frac{1}{\pi\rho}  \;,  \\
 \label{eq:sol-3}
  P^{(f)}_2  
   \A=\A 
     \frac{224}{3}
             \,\alpha\,  \frac{1}{\pi\rho} 
    \; . 
\end{eqnarray}
We note that the $x^{11}$-dependence 
of $\psi^{(A)}_{16}$ and $\psi^{(B)}_{16}$
agrees with the result of 
the gauge 5-brane\footnote{%
The $x^{11}$-dependence of 
our result coincides with that of the gauge 5-brane \cite{O1}
when we set
$\sigma_1=\sigma_2$ in the notation of \cite{O1}.
These $\sigma_1$ and $\sigma_2$ denote sizes of two SU(2) instantons
on orbifold planes at $x^{11}= 0,\pi\rho$ 
and specify the two gauge fields independently as
$ \tr {F^{(i)}}^2 \sim \Box^2\ln (1+{\sigma_i}^2/r^2)$.
}. 
We consider $\psi^{(X)}_{16}$ which remains unknown.
Since $i=24$ part of \eq{dec-1}
relates
$\psi^{(X)}_{16},\; \psi^{(C)}_{22} \;\mbox{and}\; P^{(f)}_{16}$,
it is necessary to use $i=18$ part of \eq{dec-3}
and $i=20$ part of \eq{dec-1}.
Solving these conditions,
we find that
\begin{eqnarray}
\label{eq:sol-4}
 \psi^{(X)}_{16} \A=\A \psi^{(B)}_{16} = 
    -\frac{448}{3}
 \,\alpha Q^2
  \left\{ 
   \frac{1}{2\pi\rho}(x^{11})^2 - \frac{1}{2}\,x^{11} 
    + \frac{\pi\rho}{12}
  \right\} \;, \\
 \psi^{(C)}_{10}\A = \A 0  \;, \\
\label{eq:pc-22}
 11\psi^{(C)}_{22} - 3QP^{(f)}_{16} \A =\A  36 Q\psi^{(B)}_{16} =
     -5376
 \,\alpha Q^3
  \left\{ 
   \frac{1}{2\pi\rho}(x^{11})^2 - \frac{1}{2}\,x^{11} 
    + \frac{\pi\rho}{12}
  \right\} 
\end{eqnarray}
where $P^{(f)}_{16}$ is determined by $i=30$ part of \eq{dec-3}.
$i=12$ part of \eq{dec-1}
leads to 
$P^{(f)}_4 = 0 $ .

\vs
Next, we turn to  the $x^{11}$-independent part.
By the same discussion on the $x^{11}$-dependent part,
\eq{s-3}, \eq{s-4} and \eq{s-5} yield
$ \dal A \sim \dal B \sim \dal X \sim r^{-16} $
which leads to $A \sim B \sim X \sim r^{-14}$
and $g=0$.
\eq{s-1} gives $\dal C \sim \del Y_0\, \del f$,
namely $ C \sim r^{-8}$.
Consequently, we find expansions of fields in $r$ such that
\begin{eqnarray}
\label{eq:c-2}
  A_1 = \sum^{\infty}_{i=0} P^{(A)}_{14+i} \; r^{-14-i}\; , \quad
  B_1 = \sum^{\infty}_{i=0} P^{(B)}_{14+i} \; r^{-14-i}\; , \quad
  X_1 = \sum^{\infty}_{i=0} P^{(X)}_{14+i} \; r^{-14-i}\; ,
\nn\\
  C_1 = \sum^{\infty}_{i=0} P^{(C)}_{8+i} \; r^{-8-i} 
\end{eqnarray}
where $P^{(A)}_i, P^{(B)}_i, P^{(X)}_i$ and $P^{(C)}_i$
are constants.
For expansions of equations of motion
\eq{s-1}, \eq{s-5}, \eq{s-3} and \eq{s-4},
we only need the following parts of equations.

The Maxwell equation:
\begin{eqnarray}
\label{eq:dec-5}
\A\A\quad
  \sum_i 
   \biggl[ -6(i-8)Q \Bigl\{ -2 P^{(A)}_{i-8} +6 P^{(B)}_{i-8}
   - P^{(X)}_{i-8} \Bigr\} +(i-2)(i-8)P^{(C)}_{i-2}
\nn \\
\A\A\qquad\quad
     \; 
     +6(i-8)Q P^{(f)}_{i-8}  \biggr]\: r^{-i}
      = 0 
\;, \makebox[1.5in]{} \underline{ i\geq 10}
\end{eqnarray}
\indent
The Einstein equation
\begin{eqnarray}
\label{eq:dec-6}
\A(\alpha,\beta)\,:\A \quad
\sum_i \; 
  (i-2)(i-8)\Bigl\{ -P^{(A)}_{i-2}+ P^{(X)}_{i-2} \Bigr\}
 \: r^{-i} = \frac{1}{2}\, j_1\,\alpha Q^2\: r^{-16} \,\frac{1}{\pi\rho}
\;, \qquad \underline{ i\geq 16}
  \\
\label{eq:dec-7}
\A(m,n)\,:\A \quad
 {\displaystyle \sum_i }
    \biggl[  
       36 Q^2 \Bigl\{ 6 P^{(A)}_{i-14}  
       + 3P^{(X)}_{i-14}  -3P^{(C)}_{i-14}  \Bigr\} \nn\\
\A\A \qquad\quad
       -6(i-8)Q \Bigl\{ 6 P^{(A)}_{i-8}  
       + 3P^{(X)}_{i-8}  -3P^{(C)}_{i-8}  \Bigr\} 
\nn \\
\A\A \qquad\quad
       +(i-2)(i-8)\Bigl\{ 14 P^{(A)}_{i-2} 
       +42P^{(B)}_{i-2}  +7P^{(X)}_{i-2}  \Bigr\} 
    \biggr]\: r^{-i}
\nn\\
\A\A \qquad
    = j_2\, \alpha Q^2\:r^{-16}\,\frac{1}{\pi\rho}  
\;, \makebox[2.5in]{} \underline{ i\geq 16}
\\
\label{eq:dec-8}
\A(11,11)\,:\A \quad
 {\displaystyle \sum_i} 
 \biggl[ 
  -3\cdot 36 Q^2 P^{(B)}_{i-14} + 18(i-8)Q P^{(B)}_{i-8} 
\nn\\
\A\A \qquad\quad
  + (i-2)(i-8)\Bigl\{ 
    2 P^{(A)}_{i-2} +7 P^{(B)}_{i-2} \Bigr\}
 \biggr] \: r^{-i}  
   = 0 
\;, \qquad\qquad \underline{ i\geq 16}
\end{eqnarray}
From
\eq{dec-5}, \eq{dec-6} and \eq{dec-8},
we find three conditions
\begin{eqnarray}
  P^{(C)}_8 =  -\frac{3}{4}\, Q P^{(f)}_2 \;,\quad
  P^{(X)}_{14}  =  P^{(A)}_{14} 
    + \frac{1}{224}
  \,j_1\,\alpha
  \, Q^2\,\frac{1}{\pi\rho} \;,\quad
  P^{(B)}_{14} = -\frac{2}{7}\,P^{(A)}_{14}  \;.
\end{eqnarray}
Recall that $P^{(f)}_2$ is given by \eq{sol-3}.
Substituting these conditions to \eq{dec-7}, we have
\begin{eqnarray}
\label{eq:sol-5}
 P^{(A)}_{14}\A=\A -\frac{7}{2}\,P^{(B)}_{14} = 
    -\frac{14}{3}
  \alpha Q^2
  \,\frac{1}{\pi\rho} \;, \\
\label{eq:sol-6}
 P^{(X)}_{14} \A = \A 
     \frac{4}{3}
 \alpha Q^2 \,\frac{1}{\pi\rho} \;, \\
\label{eq:sol-7}
 P^{(C)}_8 \A =\A 
              -56
  \,\alpha Q\,\frac{1}{\pi\rho}
 \; . 
\end{eqnarray}

So far, we have determined $\psi_i$ and $P_i$ asymptotically.
From the results \eq{sol-1}--\eq{sol-4} and \eq{sol-5}--\eq{sol-7},
we obtain the solution of field equations as 
the dominant forms:
\begin{eqnarray}
\label{eq:res-1}
  \begin{array}[b]{rcl@{\qquad}rcl}
\vs
{\displaystyle
 -\frac{2}{7}\,A_1}
 \A = \A  {\displaystyle
B_1  = X_1 = 
    \frac{4}{3}\,
 \alpha Q^2
  \: r^{-14}\,   \frac{1}{\pi\rho}  } \;,   \A
\A\A \\
C_1 \A =\A {\displaystyle
    -56
  \,\alpha Q\: r^{-8}\,\frac{1}{\pi\rho}  } \;,  \A
f \A = \A {\displaystyle
    \frac{224}{3}
 \,\alpha\: r^{-2}
  \frac{1}{\pi\rho} \;,} 
  \end{array} 
\end{eqnarray}
\mvs
\begin{eqnarray}
 -\frac{2}{7}\, \phi_A \A = \A   \phi_B  =  \phi_X  =
    -\frac{448}{3}
 \,\alpha Q^2 \: r^{-16}
  \left\{ 
   \frac{1}{2\pi\rho}(x^{11})^2 - \frac{1}{2}\,x^{11} 
    + \frac{\pi\rho}{12}
   \right\} \;, \nn\\
\phi_C \A=\A 
    56
 \,\alpha Q\: r^{-8}\,\frac{1}{\pi\rho} 
 \nn
\;.
\end{eqnarray}
We note that $\psi_8^{(C)}$ and $P_8^{(C)}$ are identical but 
opposite in sign. So, $\Phi_{1C}$ vanishes at this order.
Moreover, $\psi_8^{(C)}$ is independent of $x^{11}$ accidentally,
but $\psi_i^{(C)}$ for higher $i$ depends on $x^{11}$ such as 
$\psi_{22}^{(C)}$ in \eq{pc-22}.

\subsubsection{$ r^6 \ll Q $ region}

We first consider the $x^{11}$-dependent part, and
repeat the procedure similar to the one 
in $r^6 \gg Q$ region.
From \eq{rlq}, source terms are $J_1\sim J_2\sim r^{-4}$.
Suppose $\phi \sim r^p$ approximately, 
except for the $x^{11}$-dependence.
Then equations of motion lead to
$ \phi_A \sim \phi_B \sim \phi_X \sim r^2 $ and
$ \phi_C \sim f \sim r^{-2}$.
So, we find expansions of fields as
\begin{eqnarray}
\label{eq:c-3}
 \phi_A = {\displaystyle \sum^{\infty}_{i=0}} 
          \psi^{(A)}_{2+i}\;r^{2+i}\;, \quad 
 \phi_B = {\displaystyle \sum^{\infty}_{i=0}} 
          \psi^{(B)}_{2+1}\;r^{2+i} \; , \quad 
 \phi_X = {\displaystyle \sum^{\infty}_{i=0}} 
          \psi^{(X)}_{2+i}\;r^{2+i} \; , 
\nn \\
 \phi_C = {\displaystyle \sum^{\infty}_{i=0}} 
           \psi^{(C)}_{-2+i}\;r^{-2+i} \; , \quad 
 f = {\displaystyle\sum^{\infty}_{i=0}} P^{(f)}_{-2+i} \;r^{-2+i} 
\end{eqnarray}
where $ \psi^{(A)}_i, \psi^{(B)}_i, \psi^{(X)}_i$
and $\psi^{(C)}_i$ are functions of $x^{11}$,
while $P^{(f)}_i$ is a constant.
Expanding equations of motion by \eq{c-3},
we find the following results.

\indent
The Maxwell equation:
\begin{eqnarray}
\label{eq:3-max}
\sum_i \A\A
   \biggl[ 6i \Bigl\{ -2 \psi^{(A)}_{i} +6 \psi^{(B)}_{i}
   - \psi^{(X)}_{i} \Bigr\} +i(i+6)\psi^{(C)}_{i}  
     +6i P^{(f)}_{i}  \biggr]\: r^{i-2}
      = 0 
\;, \qquad \underline{i\geq -2}
\end{eqnarray}
\indent
The Einstein equation
\begin{eqnarray}
\label{eq:3-a}
\A(\alpha,\beta)\,:\A\quad
\sum_i  \biggl[ 
  i(i+6)\Bigl\{ -\psi^{(A)}_{i}+ \psi^{(X)}_{i} \Bigr\}
  + Q \Bigl\{ 
  \del^2_{11}\psi^{(A)}_{i+4} + 8 \del^2_{11}\psi^{(B)}_{i+4}
      \Bigr\}
 \biggr]\: r^{i-2} \nn\\
\A\A \qquad 
  = -\frac{1}{2}\, \tilde j_1\,\alpha \: r^{-4} \,\frac{1}{\pi\rho}
\;,  \makebox[2.7in]{} \underline{i\geq -2}
  \\
\label{eq:3-m}
\A(m,n)\,:\A \quad
 {\displaystyle \sum_i }
    \biggl[  
      (36+ 6i) \Bigl\{ 6 \psi^{(A)}_{i}  
       + 3\psi^{(X)}_{i}  -3\psi^{(C)}_{i}  \Bigr\} 
       +i(i+6)\Bigl\{ 14 \psi^{(A)}_{i} 
       +42\psi^{(B)}_{i}  +7\psi^{(X)}_{i}  \Bigr\} 
  \nn \\
  \A\A \qquad\quad 
       + Q\Bigl\{
    16\del^2_{11}\psi^{(A)}_{i+4} + 56\del^2_{11}\psi^{(B)}_{i+4}  \Bigr\}
    \biggr]\: r^{i-2}
    = -\tilde j_2\, \alpha \:r^{-4}\,\frac{1}{\pi\rho}
\;, \qquad \underline{i\geq -2}  \\
\label{eq:3-11}
\A(11,11)\,:\A\quad
 {\displaystyle \sum_i} \;
   (i+6)
  \Bigl\{ 2i\psi^{(A)}_i + (7i-18)\psi^{(B)}_i \Bigr\}
  \: r^{i-2} 
= 0 
\;,\qquad \qquad\qquad\quad \underline{i\geq 2}
\end{eqnarray}
Solving these equations, we obtain
\begin{eqnarray}
\label{eq:sol-8}
  \psi^{(A)}_2 \A = \A 
  \psi^{(B)}_2 =
     \frac{28}{3}
  \,\alpha Q^{-1}
  \left\{ 
   \frac{1}{2\pi\rho}(x^{11})^2 - \frac{1}{2}\,x^{11} 
    + \frac{\pi\rho}{12}
  \right\} \;,  \\
\label{eq:sol-9}
 \psi^{(X)}_2 \A=\A
    - \frac{560}{3} 
   \,\alpha Q^{-1}
  \left\{ 
   \frac{1}{2\pi\rho}(x^{11})^2 - \frac{1}{2}\,x^{11} 
    + \frac{\pi\rho}{12}
  \right\}  \;, \\
\label{eq:sol-10}
\psi^{(C)}_{-2} \A=\A  
 -\frac{3}{2} P^{(f)}_{-2} = 
    \frac{182}{3}
 \alpha \,\frac{1}{\pi\rho} \;,\\
 \psi^{(C)}_2 \A = \A 
   -168
 \,\alpha Q^{-1}
 \left\{ 
   \frac{1}{2\pi\rho}(x^{11})^2 - \frac{1}{2}\,x^{11} 
    + \frac{\pi\rho}{12}
  \right\} 
\; .
\end{eqnarray} 
We note that, similarly to the $r^6\gg Q$ case,
we need $\psi^{(C)}_{2}$ to obtain $\psi^{(X)}_2$.

\vs
Finally, we consider the $x^{11}$-independent part.
From equations of motion, we find 
$A_1 \sim B_1 \sim X_1 \sim C_1 \sim r^{-2}$,
and thus
\begin{eqnarray}
\label{eq:c-4}
  A_1 = \sum^{\infty}_{i=0} P^{(A)}_{-2+i} \; r^{-2+i} \;,\quad 
  B_1 = \sum^{\infty}_{i=0} P^{(B)}_{-2+i} \; r^{-2+i} \;,\quad 
  X_1 = \sum^{\infty}_{i=0} P^{(X)}_{-2+i} \; r^{-2+i} \;,
\nn\\
  C_1 = \sum^{\infty}_{i=0} P^{(C)}_{-2+i} \; r^{-2+i} 
\end{eqnarray}
where $ P^{(A)}_i, P^{(B)}_i, P^{(X)}_i$ and $P^{(C)}_i$ are constants.
Expansions of equations are as follows.

The Maxwell equation:
\begin{eqnarray}
\sum_i \A\A
   \biggl[ 6i \Bigl\{ -2 P^{(A)}_{i} +6 P^{(B)}_{i}
   - P^{(X)}_{i} \Bigr\} +i(i+6)P^{(C)}_{i}  
   - 6i P^{(f)}_{i}  \biggr]\: r^{i-2}
      = 0 
\;, \qquad \underline{i\geq -2}
\end{eqnarray}
\indent
The Einstein equation
\begin{eqnarray}
\A (\alpha,\beta)\,: \A  \qquad
\sum_i \;
  i(i+6)\Bigl\{- P^{(A)}_{i}+ P^{(X)}_{i} \Bigr\}
 \: r^{i-2} 
   = \frac{1}{2}\, \tilde j_1\,\alpha \: r^{-4} \,\frac{1}{\pi\rho}
\;, \qquad\qquad \underline{i\geq -2}
 \\
\A (m,n)\,: \A \qquad
 {\displaystyle \sum_i }\,
    \biggl[  
      (36+ 6i) \Bigl\{ 6 P^{(A)}_{i}  
       + 3P^{(X)}_{i}  -3P^{(C)}_{i}  \Bigr\} 
       +i(i+6)\Bigl\{ 14 P^{(A)}_{i} 
       +42P^{(B)}_{i}  +7P^{(X)}_{i}  \Bigr\} 
    \biggr]\: r^{i-2} \nn  \\
\A\A  \qquad\quad 
   = \tilde j_2\, \alpha \:r^{-4}\,\frac{1}{\pi\rho}  
\;, \makebox[2.5in]{} \underline{i\geq -2}
\\
\A (11,11)\,: \A \qquad
 {\displaystyle \sum_i} \;
 (i+6)
  \Bigl\{ 2iP^{(A)}_i + (7i-18)P^{(B)}_i \Bigr\}
 \: r^{i-2} 
= 0 
\;, 
\qquad\qquad \underline{i\geq -2}
\end{eqnarray}
From these equations,
we obtain 
\begin{eqnarray}
\label{eq:sol-11}
  P^{(A)}_{-2} =
  -8 P^{(B)}_{-2} =
   -\frac{14}{15}
 \,\alpha \,\frac{1}{\pi\rho}  
\;,\quad   P^{(X)}_{-2} =
    \frac{287}{30}
  \,\alpha \,\frac{1}{\pi\rho} 
\;,\quad P^{(C)}_{-2} =
    -\frac{301}{6}\,
 \alpha \,\frac{1}{\pi\rho} 
\; .
\end{eqnarray} 

Consequently,
from the above results \eq{sol-8}--\eq{sol-10} and \eq{sol-11},
the solution is given by
\begin{eqnarray}
  \begin{array}{rcl@{\qquad\quad}rcl}
\vs
  A_1 \A = \A    {\displaystyle
                -8 B_1 =
    -\frac{14}{15}
   \,\alpha \: r^{-2}\,\frac{1}{\pi\rho}  } \;,  \A
  X_1 \A=\A {\displaystyle
     \frac{287}{30}
  \,\alpha \: r^{-2}\,\frac{1}{\pi\rho} }  \;, \\
 C_1 \A=\A  {\displaystyle
   -\frac{301}{6}
 \,\alpha\: r^{-2} \,\frac{1}{\pi\rho} } \;,\A
 f \A = \A {\displaystyle
    -\frac{364}{9}
 \,\alpha \: r^{-2}\,\frac{1}{\pi\rho}  \;, }
  \end{array}\nn
\end{eqnarray}
\mvs
\begin{eqnarray}
\label{eq:res-2}
 \phi_A \A = \A \phi_B =
    \frac{28}{3}
  \,\alpha Q^{-1}\: r^2
  \left\{ 
   \frac{1}{2\pi\rho}(x^{11})^2 - \frac{1}{2}\,x^{11} 
    + \frac{\pi\rho}{12}
  \right\}  \;, \\
 \phi_X \A =\A
    - \frac{560}{3}
  \,\alpha Q^{-1} \: r^2
  \left\{ 
   \frac{1}{2\pi\rho}(x^{11})^2 - \frac{1}{2}\,x^{11} 
    + \frac{\pi\rho}{12}
  \right\} \nn \;, \\
 \phi_C \A= \A 
    \frac{182}{3}\,
 \alpha\: r^{-2}\,
   \frac{1}{\pi\rho} \nn
\;.
\end{eqnarray}

\vs\vs
In two asymptotic regions $r^6\gg Q$ and $r^6 \ll Q$,
we have found solutions \eq{res-1} and \eq{res-2}.
In an intermediate region $r^6 \sim Q$, on the other hand,
it is not clear how the solution behaves.
However,
we can solve equations of motion \eq{e-a}--\eq{m-2} numerically
by using \eq{res-1} and \eq{res-2} as boundary conditions.
In appendix, we plot $\Phi_{1A}=A_1 + \phi_A$ etc. 
as functions of $x^{11}$ and $r$.
The results become smooth surfaces, and thus it is certain that
\eq{res-1} and \eq{res-2} describe a smooth solution
of equations \eq{e-a}--\eq{m-2}.

\section{Interpretations of the solution}

We have found the M2-brane solution of heterotic M-theory
by solving 
equations of motion up to order $\kappa^{2/3}$
asymptotically in power series in $r$ or $r^{-1}$.
We discuss interpretations of this solution.

From the results of the previous section, 
we obtain the metric up to order $\kappa^{2/3}$.
From \eq{metric}, we recall that 
\begin{eqnarray}
  g_{\alpha\beta} = e^{2A}\,\eta_{\alpha\beta}
  = \exp \Bigl[ 2 \Bigl\{  \; \frac{1}{3}\,Y_0 +\kappa^{2/3}\Phi_{1A} 
     + \mbox{O}(\kappa^{4/3})\Bigr\} \Bigr] 
 \,\eta_{\alpha\beta}
\end{eqnarray}
where 
$ \Phi_{1A} = \phi_A + A_1 $ and so on.
Then, from \eq{res-1} and \eq{res-2}, we find that

\svs\noindent
$\underline{r^6 \gg Q}$
\begin{eqnarray}
  g_{\alpha\beta} \A=\A 
   \left[\;
    1+ 2\kappa^{2/3} \alpha \, Q^2 
\left\{
         \frac{1568}{3} \: r^{-16}
  \,W(x^{11})
   -    
 \frac{14}{3}
  \: r^{-14}\,   \frac{1}{\pi\rho}
\right\}  
 + \mbox{O}(\kappa^{4/3})
    \right] \eta_{\alpha\beta}
\;, \nn\\
  g_{mn} \A=\A 
   \left[\;
    1+ 2\kappa^{2/3} \alpha \, Q^2 
\left\{
         -\frac{448}{3} \: r^{-16} \,W(x^{11})
   + \frac{4}{3}
  \: r^{-14}\,   \frac{1}{\pi\rho}
\right\}  
 + \mbox{O}(\kappa^{4/3})
    \right] \delta_{mn}
\;, \\
  g_{11,11} \A=\A 
   \left[\;
    1+ 2\kappa^{2/3} \alpha \, Q^2 
\left\{
         -\frac{448}{3} \: r^{-16}\,W(x^{11})
   + \frac{4}{3}
  \: r^{-14}\,   \frac{1}{\pi\rho}
\right\}  
 + \mbox{O}(\kappa^{4/3})
    \right] 
\nn \;,
\end{eqnarray}
$\underline{r^6 \ll Q}$
\begin{eqnarray}
  g_{\alpha\beta} \A=\A r^4 Q^{-2/3}
   \left[\;
    1+ 2\kappa^{2/3} \alpha 
\left\{
         \frac{28}{3} Q^{-1}\: r^2\,W(x^{11})
   - \frac{14}{15}
    \: r^{-2}\,   \frac{1}{\pi\rho}
\right\}  
 + \mbox{O}(\kappa^{4/3})
    \right] \eta_{\alpha\beta}
\;, \nn\\  
  g_{mn} \A=\A r^{-2} Q^{1/3}
   \left[\;
    1+ 2\kappa^{2/3} \alpha 
\left\{
         \frac{28}{3} Q^{-1}\: r^2\,W(x^{11})
   +\frac{7}{60}  \: r^{-2}\,   \frac{1}{\pi\rho}
\right\}  
 + \mbox{O}(\kappa^{4/3})
    \right] \delta_{mn}
\;, \\  
  g_{11,11} \A=\A r^4 Q^{-2/3}
   \left[\;
    1+ 2\kappa^{2/3} \alpha 
\left\{
         -\frac{560}{3} Q^{-1}\: r^2\,W(x^{11})
   +\frac{287}{30}\: r^{-2}\,   \frac{1}{\pi\rho}
\right\}  
 + \mbox{O}(\kappa^{4/3})
    \right] 
\nn
\end{eqnarray}
where 
\begin{eqnarray}
W(x^{11}) \equiv
   \frac{1}{2\pi\rho}(x^{11})^2 - \frac{1}{2}\,x^{11} 
    + \frac{\pi\rho}{12}
\;.
\end{eqnarray}
From the eleven-dimensional point of view,
at order $\kappa^0$, this is the usual M2-brane solution
of eleven-dimensional supergravity, but it receives
the correction of order $\kappa^{2/3}$.
In particular, it has a non-trivial $x^{11}$-dependence
$W(x^{11})$
similar to the gauge 5-brane solution \cite{O1}.
This correction has its origin in the source terms
which consist of the Gauss-Bonnet terms in equations of motion \eq{einstein}, 
and their contributions, \eq{j1} and \eq{j2}, are given by 
the non-trivial metric \eq{metric}.
The Gauss-Bonnet terms,
on the other hand, are required by
the cancellation of gravitational anomaly and supersymmetry \cite{RW,O2}.
This anomaly is caused by
the $Z_2$ singularities of the orbifold \cite{AW,HW1,HW2}.
Therefore, the correction is regarded as a gravitational effect of
the $Z_2$ singularities.
If we are far away from the M2-brane $(r\rightarrow \infty)$, 
$\;A,B,X$ and $C$ become zero.
Thus we see that
the metric becomes flat Minkovski and the correction vanishes.

Now, we discuss the solution from the viewpoint of 
ten-dimensional string theory.
By the suggestion of Ho\v{r}ava and Witten \cite{HW1},
this solution is interpreted  as the fundamental string solution of 
strongly coupled heterotic theory at low-energy.
It is also suggested that the strong coupling limit corresponds
to large radius of the orbifold.
Then, the $x^{11}$-dependence which we have found explicitly
must be the strong coupling correction. 
To check this phenomenon, we consider the string coupling
constant.
When we compactify M-theory on $S^1 \times M^{10}$,
the string coupling constant $g_s$ and
the Regge slope $\alpha'$ are written by $\kappa$ and 
$R_{11}$, the radius of the $S^1$ \cite{W2}. 
If we consider the $S^1/Z_2$ compactification, it is necessary
to regard $R_{11}$ as the volume of $S^1/Z_2$ given by 
$\int (g_{11,11})^{1/2} \, dx^{11}$
because of the $x^{11}$-dependence, and thus we have
\begin{eqnarray}
  g_s^2 = 2 \pi^2 \left( \frac{\kappa}{4\pi} \right)^{-2/3}
          \left( \int^{\pi\rho}_0 e^X \, dx^{11} 
                     \right)^{3}
      = {\alpha'}^{-1}
          \left( \int^{\pi\rho}_0 e^X \, dx^{11}\right)^2
\;.
\end{eqnarray}
If we consider $\kappa$ as a unit, $g_s^2$ is given as,
for $r^6 \gg Q$ :
\begin{eqnarray}
  g_s^2 = 2\pi^2 \left( \frac{\kappa}{4\pi} \right)^{-2/3}
       \left[ (\pi\rho)^3 -52\alpha\, \kappa^{2/3} Q^2\, r^{-14}
              (\pi\rho)^2 \right] 
\end{eqnarray}
and for $r^6\ll Q$ :
\begin{eqnarray}
  g_s^2 = 2\pi^2 \left( \frac{\kappa}{4\pi} \right)^{-2/3}
       \left[  Q^{-1} r^6(\pi\rho)^3 
             + \frac{287}{10}\,\alpha\, \kappa^{2/3} Q^{-1} r^{4}
              (\pi\rho)^2 \right] 
\; .
\end{eqnarray}
As expected,
we see that $\rho \ll 1$ leads to weak coupling limit, while
$\rho \gg 1$ leads to strong coupling limit.

\section{Conclusion}

In this paper,
we have considered the M2-brane solution of heterotic M-theory 
with the Gauss-Bonnet $R^2$ terms up to order $\kappa^{2/3}$.
We derived equations of motion with source terms of order $\kappa^{2/3}$.
By the ansatz, only the Gauss-Bonnet terms contributed to source terms
and the equations became non-linear which required $\kappa$ expansion.
We solved them asymptotically by expanding fields in $r$.
The result is the usual M2-brane solution at order $\kappa^0$,
but it receives a correction of order $\kappa^{2/3}$
which, in particular, has the $x^{11}$-dependence 
in the same form as the gauge 5-brane solution \cite{O1}.
In appendix,
we plot the solution as a function of $x^{11}$ and $r$.
It confirms that the asymptotic solutions connected smoothly.

From the eleven-dimensional point of view,
this $\kappa^{2/3}$ correction can be regarded
as the gravitational effect of an orbifold with 
$Z_2$ singularities, because the Gauss-Bonnet terms are required 
by the cancellation of the gravitational anomaly and supersymmetry,
and only the metric contributes to the source terms
through the Gauss-Bonnet terms in our case.

We discussed 
interpretations of this solution 
as the strongly coupled fundamental heterotic string at low-energy.
Integrating this solution over $x^{11}$,
we saw the expected behavior of the string coupling constant
with a radius of the orbifold.

On the other hand, we have found the $x^{11}$-dependence explicitly 
which must have plenty of new information.
Since it describes the strong coupling effect of string theory,
it is very significant to analyze this solution with the $x^{11}$-dependence
as a background solution in strongly coupled string theory or M-theory itself.
We hope to find new applications of this solution.
In addition, it is interesting to examine 
how this solution preserves supersymmetry
with the correction of order $\kappa^{2/3}$ explicitly.

\section*{Acknowledgments}
I would like to thank Kiyoshi Higashijima and Nobuyoshi Ohta 
for useful discussions and careful reading the manuscript.

\newpage
\section*{Appendix}

We solve equations of motion \eq{e-a}--\eq{m-2} numerically
by using
\eq{res-1} and \eq{res-2} 
as boundary conditions.
For example, we choose $\rho=1,\; Q=100$ and
plot $\Phi_{1A}=A_1 + \phi_A$ etc. in \eq{warp1} and \eq{warp2} 
as functions of $x^{11}$ and $r=(x^m x^n \delta_{mn})^{1/2}$.
The range of the $x^{11}$- and $r$-axes are $[0,\pi]$ and 
$[R_0,R]$ where $R_0=(10^{-1}Q)^{1/6} \sim  1.47$ and
$R = (10\, Q)^{1/6}\sim 3.16\, $.
These figures indicate that 
the asymptotic solutions, \eq{res-1} and \eq{res-2},
are connected smoothly.

\vs\vs
\begin{figure}[h]

\centerline{
\begin{minipage}[b]{4cm}
\centerline{\epsfbox{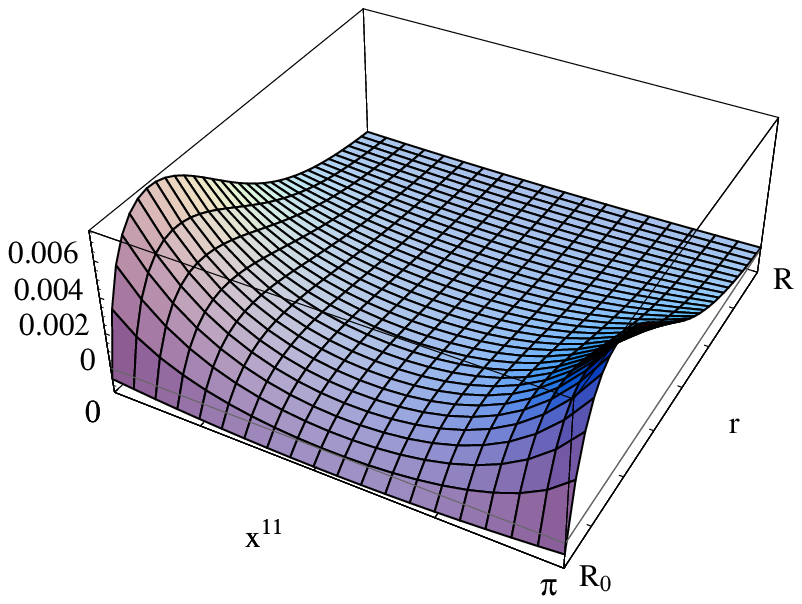}}
   \caption{$\Phi_{1A}$}
\end{minipage}
\hspace{1.7in}
\begin{minipage}[b]{4cm}
\centerline{\epsfbox{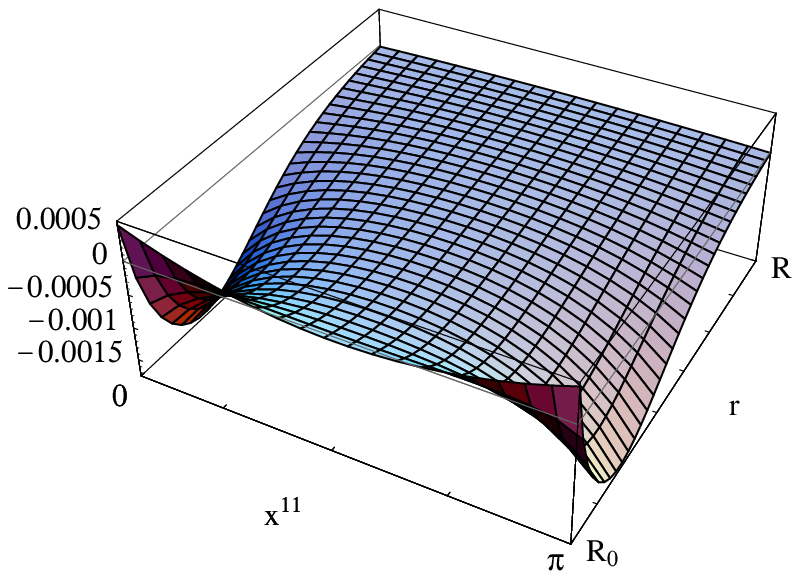}}
   \caption{$\Phi_{1B}$}
\end{minipage}
}

\Vs{8}
\centerline{
\begin{minipage}[b]{4cm}
\centerline{\epsfbox{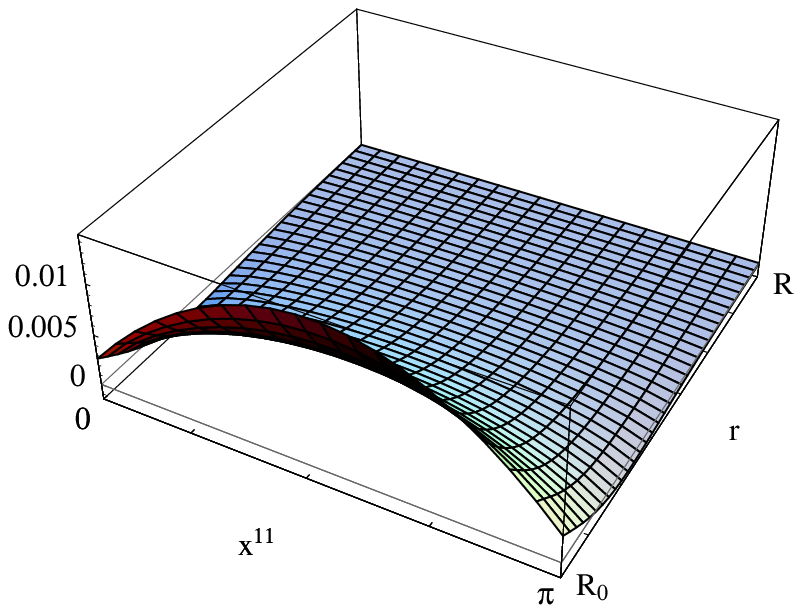}}
   \caption{$\Phi_{1X}$}
\end{minipage}
\hspace{1.7in}
\begin{minipage}[b]{4cm}
\centerline{\epsfbox{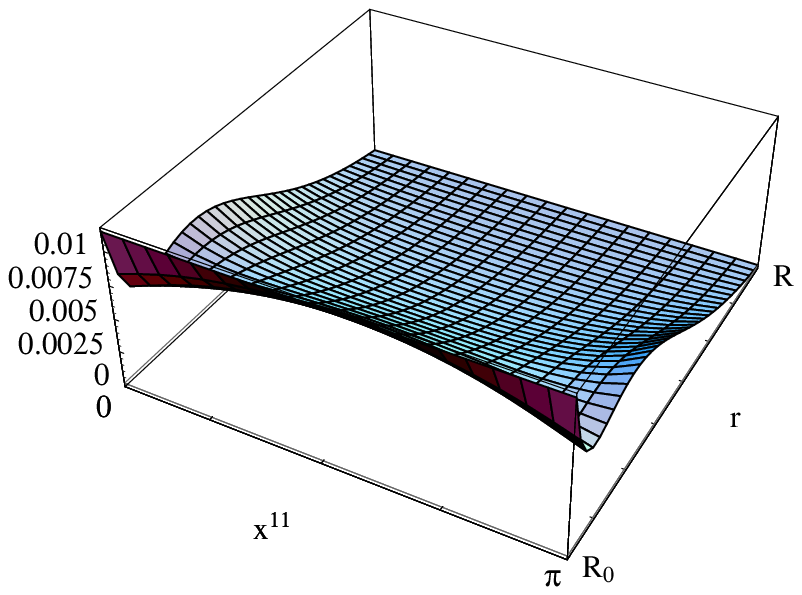}}
   \caption{$\Phi_{1C}$}
\end{minipage}
}

\vs\vs\vs
\centerline{\it
The $\kappa^{2/3}$ correction from the Gauss-Bonnet terms
}
\end{figure}

\end{document}